\documentclass[aps,pra,onecolumn,superscriptaddress,floatfix,longbibliography]{revtex4-2}
\usepackage{bm,amsmath,amsfonts,amssymb,braket}
\usepackage{latexsym,dsfont,array,layout,mathrsfs,textcase}
\usepackage{graphicx,graphics,subfigure,xcolor}
\usepackage{comment,verbatim}
\usepackage{float}
\usepackage{times}
\usepackage{xspace}
\usepackage{stmaryrd}
\usepackage{multirow}
\usepackage{mathtools}
\usepackage{booktabs}
\usepackage{appendix}
\usepackage[colorlinks,linkcolor=blue,citecolor=blue,urlcolor=blue]{hyperref}
\usepackage{epstopdf}

\begin{document}

\title{Exact solution of a two-parameter extended Bariev model}

\author{Mingchen Zheng}
\affiliation{Beijing National Laboratory for Condensed Matter Physics, Institute of
Physics, Chinese Academy of Sciences, Beijing 100190, China}
\affiliation{School of Physical Sciences, University of Chinese Academy of Sciences,
Beijing 100049, China}
\author{Xin Zhang}
\affiliation{Beijing National Laboratory for Condensed Matter Physics, Institute of
Physics, Chinese Academy of Sciences, Beijing 100190, China}
\author{Junpeng Cao}
\affiliation{Beijing National Laboratory for Condensed Matter Physics, Institute of
Physics, Chinese Academy of Sciences, Beijing 100190, China}
\affiliation{School of Physical Sciences, University of Chinese Academy of Sciences,
Beijing 100049, China}
\affiliation{Songshan lake Materials Laboratory, Dongguan, Guangdong 523808, China}
\affiliation{Peng Huanwu Center for Fundamental Theory, Xi'an 710127, China}
\author{Wen-Li Yang}
\affiliation{Peng Huanwu Center for Fundamental Theory, Xi'an 710127, China}
\affiliation{Institute of Modern Physics, Northwest University, Xi'an 710127, China}
\affiliation{Shaanxi Key Laboratory for Theoretical Physics Frontiers, Xi'an 710127, China}
\author{Yupeng Wang}
\email{yupeng@iphy.ac.cn}
\affiliation{Beijing National Laboratory for Condensed Matter Physics, Institute of
Physics, Chinese Academy of Sciences, Beijing 100190, China}

\date{\today}

\begin{abstract}

An exactly solvable strongly correlated electron model with two independent parameters is constructed in the frame of the quantum inverse scattering method, which can be seen as a generalization of the Bariev model. Through the Bethe ansatz method, a set of Bethe ansatz equations is derived. In the thermodynamic limit, to study the ground state of the model, we obtain the integral equations for the density of Bethe roots. Numerical validation are done to confirm the accuracy of our analytic results. 
   
\end{abstract}

\maketitle

\section{Introduction}
In the field of strongly correlated electron systems, persistent efforts have led to noteworthy progress in the investigation of novel integrable models. 
Many interesting and widely applicable one-dimension models have been demonstrated to be integrable. 
Among them, remarkable ones are the one-dimensional Hubbard model \cite{Lieb1,Essler1,Li,Hou}, the supersymmetric $t-J$ model \cite{FC1,Schlottmann1,Essler2,Zhang} and the Bariev model \cite{Bariev1,Bariev2,Bariev3,Bariev4}.  
Based on the exact solutions derived from
the integrability theory, many properties of these models have been extracted \cite{Essler1}. 
Generally, integrable models can serve as valuable benchmarks, making the exploration for novel integrable models crucial. 

One essential approach for constructing integrable models is the quantum inverse scattering method \cite{Faddeev1,Sklyanin1,ABA,Wadati1}, 
which starts from the $R$ matrix and the  Yang-Baxter equation and gives rise to meaningful integrable conserved quantities.  
In 1968, Lieb and Wu provided exact solution of the Hubbard model \cite{Lieb1}, and later, 
Shastry made a remarkable contribution which provides the $R$ matrix for the Hubbard model with two parameters \cite{Shastry1}.  
The 1D Bariev model with periodic boundary conditions was proposed and solved by Bariev \cite{Bariev1}. 
Bariev, Alcaraz, et al. continued their research and found several new integrable models \cite{Bariev5,Alcaraz1,Alcaraz2}. 
The integrability of Bariev model was studied by Zhou \cite{Zhou1}. 
Similarly, the $R$ matrix of the Bareiv model is a non-additive spectral two-parameter matrix. There is only one parameter in Bariev model. The presence of more parameters in the $R$ matrix suggests the possibility of finding an extended Hamiltonian, 
including all parameters appeared in the $R$ matrix. 

Based on the Yang-Baxter algebraic structure of the Bareiv model, we derive a two-parameter integrable model, which we refer to as the extended Bariev model. Interestingly, we later discovered that this model coincides with an integrable model previously proposed by Zhang in 2020 \cite{Zhang1}, derived using a different 
$R$ matrix. However, the exact solutions of the model are not derived in Ref. \cite{Zhang1}. In this work we presents the Bethe ansatz equations of this model and calculates the ground state energy in the thermodynamic limit. In addition, a one-parameter integrable model beyond Bariev model is also studied by taking a proper limit. 

The organization of the paper is as follows: In section II, we construct an integrable extended Bariev model, based on the quantum inverse scattering method. In Section III, we demonstrate the exact solution of the model. Section IV is devoted to analysis the ground state of the model in the thermodynamic limit. 
Finally, in Section V, we focus on the degenerate single-parameter model. 



\section{Extended Bariev model and its integrability}

\begin{figure}[tp]
    \includegraphics[width=0.6\textwidth]{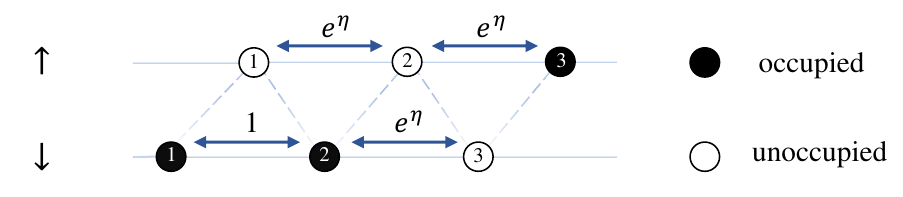} 
    \caption{Schematic diagram of the Bariev model.}
    \label{fig1}
\end{figure}

The Hamiltonian of the Bariev model under periodic boundary conditions reads \cite{Bariev1}
\begin{equation}\label{HB}
    H=-\sum_{j=1}^N \bigg[\left(c_{j\uparrow}^\dagger c_{j+1\uparrow}+c_{j+1\uparrow}^\dagger c_{j\uparrow}\right)\exp{(\eta n_{j \downarrow})}+\left(c_{j\downarrow}^\dagger c_{j+1\downarrow}+c_{j+1\downarrow}^\dagger c_{j\downarrow}\right)\exp{(\eta n_{j+1 \uparrow})}\bigg],
\end{equation}
where $c_{j \alpha}^{\dagger}$ and $c_{j \alpha}$ denote the fermionic creation and annihilation operators with spin $\alpha$ ($\uparrow $ or $\downarrow$) at site $j$ and $n_{j\alpha}$ is the number operator. The periodic boundary condition implies $c_{N+1,\alpha}=c_{1,\alpha}$ and $c^\dagger_{N+1,\alpha}=c^\dagger_{1,\alpha}$.
This model can be regarded as two coupled $XY$ chains, with the movement of electrons between adjacent lattice sites dependent on the occupancy of corresponding sites on the other chain (see Fig.\ref{fig1}). 


The $R$ matrix associated with the Bariev model is given by Zhou \cite{Zhou1} 
\begin{equation}
R_{1,2}(\lambda ,\mu )=
\addtocounter{MaxMatrixCols}{16}
\begin{pmatrix}\label{R_Matrix}
\rho_1&0&0&0&0&0&0&0&0&0&0&0&0&0&0&0\\
0&\rho _2&0&0&\rho _3&0&0&0&0&0&0&0&0&0&0&0\\
0&0&\rho _2&0&0&0&0&0&\rho _3&0&0&0&0&0&0&0\\
0&0&0&\rho _4&0&0&\rho _5&0&0&\rho _6&0&0&\rho _9&0&0&0\\
0&\rho _3&0&0&\rho _2&0&0&0&0&0&0&0&0&0&0&0\\
0&0&0&0&0&\rho _1&0&0&0&0&0&0&0&0&0&0\\
0&0&0&\rho _{12}&0&0&\rho _7&0&0&\rho _{15}&0&0&\rho _5&0&0&0\\
0&0&0&0&0&0&0&\rho _8&0&0&0&0&0&\rho _{11}&0&0\\
0&0&\rho _3&0&0&0&0&0&\rho _2&0&0&0&0&0&0&0\\
0&0&0&\rho _{13}&0&0&\rho _{15}&0&0&\rho _{10}&0&0&\rho _{6}&0&0&0\\
0&0&0&0&0&0&0&0&0&0&\rho _1&0&0&0&0&0\\
0&0&0&0&0&0&0&0&0&0&0&\rho _8&0&0&\rho _{11}&0\\
0&0&0&\rho _{14}&0&0&\rho _{12}&0&0&\rho _{13}&0&0&\rho _{4}&0&0&0\\
0&0&0&0&0&0&0&\rho _{11}&0&0&0&0&0&\rho _8&0&0\\
0&0&0&0&0&0&0&0&0&0&0&\rho _{11}&0&0&\rho _8&0\\
0&0&0&0&0&0&0&0&0&0&0&0&0&0&0&\rho _1
\end{pmatrix},
\end{equation}
with
\begin{align*} 
\rho_1&=1,\quad
\rho_2=\frac{\sqrt{1+h^2\lambda^2}\sqrt{1+h^2\mu^2}}{1+\lambda h^2 \mu},\quad 
\rho_3=\frac{(\lambda -\mu )h}{1+\lambda h^2 \mu},\\ 
\rho_4&=\frac{\sqrt{1+h^2\lambda^2}\sqrt{1+h^2\mu^2}\sqrt{1+\lambda^2}\sqrt{1+\mu^2}}{(1+\lambda \mu)(1+\lambda h^2 \mu)},\quad
\rho_5=\frac{h\sqrt{1+h^2\mu^2}\sqrt{1+\lambda^2}(\lambda -\mu )}{(1+\lambda \mu)(1+\lambda h^2 \mu)},\\
\rho_6&=\frac{\sqrt{1+h^2\mu^2}\sqrt{1+\lambda^2}(\lambda -\mu )}{(1+\lambda \mu)(1+\lambda h^2 \mu)},\quad
\rho_7=\frac{1+\lambda^2 h^2 +\lambda \mu -\lambda h^2 \mu +h^2 \mu^2+\lambda^2 h^2 \mu^2}{(1+\lambda \mu)(1+\lambda h^2 \mu)},\\
\rho_8&=\frac{\sqrt{1+\lambda ^2}\sqrt{1+\mu^2}}{1+\lambda \mu},\quad 
\rho_9=\frac{(\lambda-\mu)(\lambda -h^2\mu )}{(1+\lambda \mu)(1+\lambda h^2 \mu)},\\
\rho_{10}&=\frac{1+\lambda^2 -\lambda \mu +\lambda h^2 \mu + \mu^2+\lambda^2 h^2 \mu^2}{(1+\lambda \mu)(1+\lambda h^2 \mu)},\quad 
\rho_{11}=\frac{\lambda -\mu }{1+\lambda \mu },\\
\rho_{12}&=\frac{h\sqrt{1+h^2\lambda ^2}\sqrt{1+\mu^2}(\lambda -\mu )}{(1+\lambda \mu)(1+\lambda h^2 \mu)},\quad 
\rho_{13}=\frac{\sqrt{1+h^2\lambda ^2}\sqrt{1+\mu^2}(\lambda -\mu )}{(1+\lambda \mu)(1+\lambda h^2 \mu)},\\
\rho_{14}&=\frac{(\lambda -\mu )(h^2\lambda -\mu )}{(1+\lambda \mu )(1+\lambda h^2 \mu )},\quad 
\rho_{15}=\frac{h(\lambda -\mu )^2}{(1+\lambda \mu )(1+\lambda  h^2 \mu )},
\end{align*}
where the subscripts $\{1,2\}$ implies that $R$ matrix acts in the tensor space $V_1 \otimes V_2$ and $h=e^{\eta}$.

The $R$  matrix satisfies several relations:

(1) Initial Condition
\begin{equation}
    R_{1,2}(\lambda,\lambda)= P_{1,2},
\end{equation}
where $P_{1,2}$ is the permutation operator. 

(2) Unitary Relation
\begin{equation}
    R_{1,2}(\lambda,\mu)R_{2,1}(\mu,\lambda)= I.
\end{equation}
We can verify that the $R$ matrix (\ref{R_Matrix}) satisfies the Yang-Baxter equation \cite{Zhou1}
\begin{equation}
    R_{1,2}(\lambda,\mu)R_{1,3}(\lambda,\theta)R_{2,3}(\mu,\theta)=R_{2,3}(\mu,\theta)R_{1,3}(\lambda,\theta)R_{1,2}(\lambda,\mu).
\end{equation}
The monodromy matrix can be defined as
\begin{equation}
T_0(u,\{\theta_m\})=R_{N,0}(u,\theta _N)\dots R_{1,0}(u,\theta_1), 
\end{equation}
with $\{\theta_m|m=1,...,N\}$ being generic site-dependent inhomogeneity parameters. It can be proved that the monodromy matrix satisfies the $RTT$ relation
\begin{equation}
    R_{1,2}(\lambda ,\mu)T_1(\lambda )T_2(\mu)=T_2(\mu)T_1(\lambda )R_{1,2}(\lambda,\mu).
\end{equation} 
The transfer matrix is defined as \cite{ABA}
\begin{equation}
    t(u)=tr_0 \{T_0(u,\{\theta_m\})\}.  
\end{equation}
The transfer matrices with different spectral parameters commute with each other: $[t(u),\,t(v)]=0$ \cite{ABA}. The logarithmic derivative of $t(u)$ gives the Hamiltonian of an extended Bariev model
\begin{equation}\label{Ham1}
 \begin{split}
H=&\frac{\partial}{\partial u}{\ln t(u,\{\theta_m\})}\bigg|_{u=\theta,\{\theta _m=\theta\}}\\
=&\sum_{j=1}^N \left\{\left(\frac{1}{4\sqrt{\theta^2+1}\sqrt{h^2\theta^2+1}}+\frac{1}{4(\theta^2+1)}+\frac{h}{4\sqrt{\theta^2+1}\sqrt{h^2\theta^2+1}}+\frac{h}{4(h^2\theta^2+1)}\right)\right.\\
&\left.\times\left(\sigma _j^+\sigma _{j+1}^-\right.+\sigma _j^-\sigma _{j+1}^++\tau  _j^+\tau  _{j+1}^-+\tau  _j^-\tau  _{j+1}^+\right)\\
&+\left(-\frac{1}{4\sqrt{\theta^2+1}\sqrt{h^2\theta^2+1}}+\frac{1}{4(\theta^2+1)}-\frac{h}{4\sqrt{\theta^2+1}\sqrt{h^2\theta^2+1}}+\frac{h}{4(h^2\theta^2+1)}\right)\\
&\left.\times \left(\sigma _j^z\sigma _{j+1}^z \right.(\tau  _j^+\tau  _{j+1}^-+\tau  _j^-\tau  _{j+1}^+)+(\sigma _j^+\sigma _{j+1}^-+\sigma _j^-\sigma _{j+1}^+)\tau_j^z\tau_{j+1}^z\right)\\
&+\left(-\frac{1}{4\sqrt{\theta^2+1}\sqrt{h^2\theta^2+1}}-\frac{1}{4(\theta^2+1)}+\frac{h}{4\sqrt{\theta^2+1}\sqrt{h^2\theta^2+1}}+\frac{h}{4(h^2\theta^2+1)}\right)\\
&\left.\times \left(\sigma_j^z\right.(\tau  _j^+\tau  _{j+1}^-+\tau  _j^-\tau  _{j+1}^+)+(\sigma_j^+\sigma_{j+1}^-+\sigma_j^-\sigma_{j+1}^+)\tau_{j+1}^z\right)\\
&+\left(\frac{1}{4\sqrt{\theta^2+1}\sqrt{h^2\theta^2+1}}-\frac{1}{4(\theta^2+1)}-\frac{h}{4\sqrt{\theta^2+1}\sqrt{h^2\theta^2+1}}+\frac{h}{4(h^2\theta^2+1)}\right)\\
&\left.\times \left(\sigma_{j+1}^z\right.(\tau  _j^+\tau  _{j+1}^-+\tau  _j^-\tau  _{j+1}^+)+(\sigma_j^+\sigma_{j+1}^-+\sigma_j^-\sigma_{j+1}^+)\tau_{j}^z\right)\\
&\left.+\frac{ \theta (h^2-1)}{(\theta^2+1)(h^2\theta^2+1)}\left(\sigma_j^- \sigma_{j+1}^+ \tau_j^- \tau_{j+1}^+-\sigma_j^+ \sigma_{j+1}^- \tau_j^+ \tau_{j+1}^-  \right)\right\}.
\end{split}
\end{equation}
Here $\{\sigma_j^\alpha\}$ and $\{\tau_j^\alpha\}$ are two sets of Pauli matrices at site $j$, which commute with each other. 
By using the Jordan-Wigner transformation and multiplying a factor $(\theta^2+1)$, the Hamiltonian (\ref{Ham1}) becomes

\begin{equation}\label{H1}
    \begin{split}
H=&-\sum_{j=1}^N\left\{ \left(c_{j\downarrow}^\dagger c_{j+1\downarrow}+c_{j+1\downarrow}^\dagger c_{j\downarrow}\right)\left( 1+ \left(-\frac{\sqrt{\theta^2+1}}{\sqrt{h^2\theta^2+1}}\right.
+1-\frac{h\sqrt{\theta^2+1}}{\sqrt{h^2\theta^2+1}}+\frac{h(\theta^2+1)}{(h^2\theta^2+1)}\right)n_{j\uparrow}n_{j+1\uparrow}\right.\\
& \left.+\left(\frac{h\sqrt{\theta^2+1}}{\sqrt{h^2\theta^2+1}}-1\right)n_{j\uparrow}
+\left(\frac{\sqrt{\theta^2+1}}{\sqrt{h^2\theta^2+1}}  -1\right) n_{j+1\uparrow}\right)\\
&+\left(c_{j\uparrow}^\dagger c_{j+1\uparrow}+c_{j+1\uparrow}^\dagger c_{j\uparrow}\right)\left( 1+ \left(-\frac{\sqrt{\theta^2+1}}{\sqrt{h^2\theta^2+1}}
+1-\frac{h\sqrt{\theta^2+1}}{\sqrt{h^2\theta^2+1}}+\frac{h(\theta^2+1)}{(h^2\theta^2+1)}\right)n_{j\downarrow}n_{j+1\downarrow}\right.\\
&\left.+\left(\frac{h\sqrt{\theta^2+1}}{\sqrt{h^2\theta^2+1}}-1\right)n_{j+1\downarrow}
+\left(\frac{\sqrt{\theta^2+1}}{\sqrt{h^2\theta^2+1}}-1\right) n_{j\downarrow}\right)\\
&\left.+\frac{ \theta (h^2-1)}{(h^2\theta^2+1)}
\left( c_{j+1\uparrow}^{\dagger}c_{j\uparrow}c_{j+1\downarrow}^{\dagger}c_{j\downarrow}-c_{j\uparrow}^{\dagger}c_{j+1\uparrow}c_{j\downarrow}^{\dagger}c_{j+1\downarrow}\right)
\right\}.
\end{split}
\end{equation}


An integrable extended Bariev model is thus constructed. Compared with the conventional Bariev model (\ref{HB}), the extended one in (\ref{H1}) contains more interaction terms, e.g., the pair hopping term and the four-particle interactions between adjacent sites. We see that there are two independent parameters $\theta$ and $h$ (or $\eta$) in the Hamiltonian. When $\theta=0$, our extended Bariev model reduces to the conventional one.
It should be pointed out that the Hamiltonian (\ref{H1}) has been derived by Zhang with a completely different $R$ matrix \cite{Zhang1}. 

To ensure the Hermiticity of (\ref{H1}), for real $\eta$, ${\rm Re}[\theta]$ should be zero. Specifically, the Hermitian condition specifies four parameter regimes
\begin{align}\label{hermiticity}
\begin{aligned}
\mbox{for} \,\,\eta<0, \quad |\theta|<1 \,\,\mbox{or} \,\,|\theta|>1/h,\\
\mbox{for} \,\,\eta>0, \quad |\theta|>1\,\, \mbox{or}\,\, |\theta|<1/h.
\end{aligned}
\end{align}

\section{Bethe ansatz Solution}
In this section, we will use the Bethe ansatz method to solve the eigenvalue problem of the extended Bariev model \cite{Lieb1,Wang1}.
Since the particle number is a conserved charge in the extended Bariev model, eigenstates of the model can be constructed as
\begin{equation}
    \ket{\Psi} =\sum_{j=1}^M\sum_{\alpha_j=\uparrow,\downarrow}\sum_{x_j=1}^N\psi^{\{\alpha\}} (x_1,\dots,x_M)c_{x_1,\alpha_1}^\dagger \dots c_{x_M,\alpha_M}^\dagger\ket{0}, 
\end{equation}
where the integer $M\in[0,2N]$ is the number of electrons, $\{\alpha\}=\{\alpha_1,\dots,\alpha_m\}$ and $\ket{0}$ is the vacuum state.
One can propose the following ansatz for the wave function
\begin{equation}
    \psi^{\{\alpha\}} (x_1,\dots,x_M)=\sum_{p,q}A_p^{\{\alpha\}}(q)\exp\left(i\sum_{j=1}^M k_{p_j}x_{q_j}\right)\theta(x_{q_1}\leq x_{q_2}\leq\dots\leq x_{q_M}), 
\end{equation}
where $p=\{p_1,\dots,p_M\}$ and $q=\{q_1,\dots,q_M\}$ are the permutations of $\{1,\dots,M\}$ 
and $\theta(x_{q_1}\leq x_{q_2}\leq\dots\leq x_{q_M})$ is the generalized step function, which equals to one when the variables $x_{q_1}, x_{q_2}, \dots, x_{q_M}$ are in non-decreasing order and zero otherwise.

For the case $x_{q_j}\neq x_{q_l}$ and $x_{q_j}\neq x_{q_l}\pm 1$, the eigen-equation $H\ket{\Psi}=E\ket{\Psi}$ gives the energy
\begin{equation}
    E=-2\sum_{j=1}^M\cos k_j.
\end{equation}

Considering the case where $x_{q_j}= x_{q_{j+1}}$ or $x_{q_j} = x_{q_{j+1}}\pm 1$, we can get that the amplitude of the wave function satisfies the following equation
\begin{equation}
    A_p^{\{\alpha\}}(q)=S_{p_{j+1},p_j}(k_{p_{j+1}},k_{p_j})A_{p'}^{\{\alpha'\}}(q'), \label{S_Matrix}
\end{equation}
where $\{\alpha'\}=\{\dots,\alpha_{p_{j+1}},\alpha_{p_j},\dots,\}$, $p'=\{\dots,p_{j+1},p_{j},\dots \}$ and $q'=\{\dots,q_{j+1},q_j,\dots\}$. 

The two-body scattering matrix $S$ in (\ref{S_Matrix}) can be written in matrix form as (more details are shown in Appendix \ref{appendix:a})

\begin{equation}\label{SSSSmatrix}
    S_{12}(k_1,k_2)=
    \begin{pmatrix}
        1&0&0&0\\
        0&S_{22}
        &S_{23}&0\\
        0&S_{32}&S_{33}&0\\
        0&0&0&1
    \end{pmatrix},
\end{equation}
where 
\begin{equation}
    \begin{split}
        &S_{22}=S_{33}=\frac{e^{\eta } \left(\theta ^2+1\right) \left(e^{i k_1}-e^{i k_2}\right)}{\theta ^2 \left(e^{2 \eta +i k_1}-e^{i k_2}\right)+\left(e^{2 \eta }-1\right) \theta  \left(e^{i \left(k_1+k_2\right)}-1\right)+\left(e^{i k_1}-e^{2 \eta +i k_2}\right)},\\
&S_{23}=\frac{\left(e^{2 \eta }-1\right) \left(e^{i k_1}+\theta \right) \left(\theta  e^{i k_2}-1\right)}{\theta ^2 \left(e^{2 \eta +i k_1}-e^{i k_2}\right)+\left(e^{2 \eta }-1\right) \theta  \left(e^{i \left(k_1+k_2\right)}-1\right)+\left(e^{i k_1}-e^{2 \eta +i k_2}\right)},\\
&S_{32}=\frac{\left(e^{2 \eta }-1\right) \left(e^{i k_2}+\theta \right) \left(\theta  e^{i k_1}-1\right)}{\theta ^2 \left(e^{2 \eta +i k_1}-e^{i k_2}\right)+\left(e^{2 \eta }-1\right) \theta  \left(e^{i \left(k_1+k_2\right)}-1\right)+\left(e^{i k_1}-e^{2 \eta +i k_2}\right)}.
    \end{split}
\end{equation}

By introducing the following parameter transformation
\begin{equation}
    e^{i k_j}=\frac{\theta +e^{i \lambda _j}}{\theta  e^{i \lambda _j}-1},
\end{equation}
we can rewrite the scattering matrix in terms of $\lambda$
\begin{equation}
    \mathcal{S}_{12}(\lambda_1,\lambda_2)=\left(
\begin{array}{cccc}
 1 & 0 & 0 & 0 \\
 0 & \frac{\sin \left(\frac{1}{2} \left(\lambda _1-\lambda _2\right)\right)}{\sin \left(\frac{1}{2} \left(\lambda _1-\lambda _2\right)+i \eta \right)} & \frac{i e^{\frac{i}{2} \left(\lambda _1-\lambda _2\right)} \sinh \eta }{\sin \left(\frac{1}{2} \left(\lambda _1-\lambda _2\right)+i \eta \right)} & 0 \\
 0 & \frac{i e^{-\frac{i}{2}  \left(\lambda _1-\lambda _2\right)} \sinh\eta}{\sin \left(\frac{1}{2} \left(\lambda _1-\lambda _2\right)+i \eta \right)} & \frac{\sin \left(\frac{1}{2} \left(\lambda _1-\lambda _2\right)\right)}{\sin \left(\frac{1}{2} \left(\lambda _1-\lambda _2\right)+i \eta \right)} & 0 \\
 0 & 0 & 0 & 1 \\
\end{array}
\right). 
\end{equation}
For quantum integrable models, the many-body scattering can be expressed as the product of two-body scattering. From the periodic boundary condition $\psi(\dots,x_j+N,\dots)=\psi(\dots,x_j,\dots)$, we derive the following equations
\begin{equation}\label{Eigen1}
    \begin{split}
    A_p^{\{\alpha\}}(q)=&S_{j,j+1}(k_j,k_{j+1})S_{j,j+2}(k_j,k_{j+2})\dots S_{j,M}(k_j,k_M)\\
    &\times S_{j,1}(k_j,k_1)\dots S_{j,j-1}(k_j,k_{j-1})e^{-i k_j N} A_p^{\{\alpha\}}(q), \quad j=1,\ldots,M,
    \end{split}
\end{equation}
or equivalently
\begin{align}\label{SSe}
   A_p^{\{\alpha\}}(q)=&\left(\frac{\theta +e^{i \lambda _j}}{\theta  e^{i \lambda _j}-1}\right)^{-N} \mathcal{S}_{j,j+1}(\lambda_j,\lambda_{j+1})\mathcal{S}_{j,j+2}(\lambda_j,\lambda_{j+2})\dots \mathcal{S}_{j,M}(\lambda_j,\lambda_M)\nonumber\\
&\times \mathcal{S}_{j,1}(\lambda_j,\lambda_1)\dots \mathcal{S}_{j,j-1}(\lambda_j,\lambda_{j-1})A_p^{\{\alpha\}}(q),\quad j=1,\ldots,M.
\end{align}

The eigenvalue equation (\ref{Eigen1}) (or (\ref{SSe})) can be solved using the algebraic Bethe ansatz method (see Appendix \ref{appendix:b}), and a set of Bethe ansatz equations (BAEs) is formulated
\begin{equation}\label{BAEtr1}
    \begin{split}
        &\left[\frac{\sin \left(\frac{1}{2} \left(K_j+i \xi \right)\right)}{ \sin \left(\frac{1}{2} \left(K_j-i \xi \right)\right)}\right]^N=i^N\prod _{\alpha =1}^{\bar{M}} \frac{\sin \left(\frac{1}{2} \left(K_j-\gamma _{\alpha }-i \eta\right)\right)}{\sin \left(\frac{1}{2} \left(K_j-\gamma _{\alpha }+i \eta \right)\right)},\ \ j=1,...,M,\\
        &\prod _{j=1}^M \frac{\sin \left(\frac{1}{2} \left(K_j-\gamma_{\alpha }-i \eta \right)\right)}{\sin \left(\frac{1}{2} \left(K_j-\gamma_{\alpha }+i \eta \right)\right)}=-\prod _{\beta =1}^{\bar{M}} \frac{\sin \left(\frac{1}{2} \left(\gamma _{\alpha }-\gamma _{\beta }\right)+i \eta \right)}{\sin \left(\frac{1}{2} \left(\gamma _{\alpha }-\gamma _{\beta }\right)-i \eta \right)}, \ \ \alpha=1,...,\bar{M},  
    \end{split}
    \end{equation}
where $\theta=i e^{\xi}$, $K_j=\lambda_j+\frac{\pi}{2}$. Here, $\{\gamma_\alpha\}$ are the introduced spin rapidities and $\bar{M}\in[0,M]$ denotes the number of particles with spin up or spin down. 
The corresponding eigenvalue of the Hamiltonian (\ref{H1}) is 
\begin{equation}\label{energyB}
    E=-\sum_{j=1}^M \frac{2 \sinh \xi \, \sin K_j}{\cosh \xi -\cos K_j}. 
\end{equation}
The equations \eqref{BAEtr1} and \eqref{energyB} fully determine the spectrum of Hamiltonian \eqref{H1}. For small-scale systems, we have checked numerically that the spectrum obtained from equations \eqref{BAEtr1} and \eqref{energyB} is consistent with the one given by the exact diagonalization of the Hamiltonian.

\section{The ground state}

When $N$ is even, under the substitution \(\{\theta, \eta\}=\{-1/\theta, -\eta\}\), we find the following one-to-one correspondence for the BAEs and the energy
\begin{align}
\xi\to -\xi,\quad K_j\to -K_j,\quad \gamma_\alpha \to -\gamma_\alpha,\quad E\to E.\label{correspondence}
\end{align}
Due to the symmetry in (\ref{correspondence}), we only analyze the ground state properties in two Hermitian cases, i.e., 
$$\eta < 0,\,\,0 <\text{Im}[\theta] < 1\,\,\mbox{and}\,\, \eta > 0,\,\,0<\text{Im}[\theta] < 1/h.$$


\subsection{\(\eta < 0\) and \(0 <\text{Im}[\theta] < 1\)}

\begin{figure}[h]
    \centering
    \includegraphics[width=0.65\textwidth]{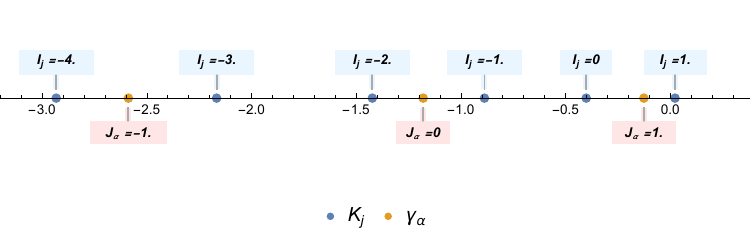}
    \caption{Numerical result of Bethe roots corresponding to the ground state where \(N=6\), \(M=6\), \(\bar{M}=3\), \(\eta =-0.5\) and \(\theta =0.25 i\). }
    \label{figBie1}
\end{figure}

Assuming that both $N$ and $M$ are even, then the total magnetization of the ground state of the system is zero, i.e., $\bar{M}=M/2$. The Bethe roots $K_j$ and $\gamma_\alpha$ corresponding to the ground state are all distributed on the real axis (see Fig.\ref{figBie1}). Taking the logarithm of the BAEs \eqref{BAEtr1}, we obtain
\begin{equation}
    \begin{split}\label{BAEtr2}
    &- N \Theta \left(K_j,\frac{\xi }{2}\right)-\sum _{\alpha =1}^{\bar{M}} \Theta \left(K_j-\gamma _{\alpha },\frac{\eta }{2}\right)=2 \pi  I_j, \ \ j=1,...,M,\\
    &-\sum _{j=1}^M \Theta \left(\gamma _{\alpha }-K_j,\frac{\eta }{2}\right)+\sum _{\beta =1}^{\bar{M}} \Theta \left(\gamma _{\alpha }-\gamma _{\beta },\eta \right)=2 \pi  J_{\alpha }, \ \ \alpha=1,...,\bar{M},
    \end{split}
    \end{equation}
where $\Theta  (k,\alpha )=2 \arctan\left(\coth (\alpha ) \tan \left(\frac{k}{2}\right)\right)$. 

Each set of $\{I_j,\ J_\alpha|j=1,...,M,\quad \alpha=1,...,\bar{M}\}$ represents a specific eigenstate of the Hamiltonian. After some calculations, we find numerically that the ground state can be obtained with 
\begin{equation}
    \begin{split}
        I_j&=j-(M+4)/2 \quad \text{for} \quad j=1,...,M, \\
        J_\alpha&=\alpha -(\bar{M}+1)/2 \quad \text{for} \quad \alpha=1,...,\bar{M}. 
    \end{split}
\end{equation} 
We assume that in the thermodynamic limit where $N$, $M$, and $\bar M$ tend to $\infty$, while the ratios $M/N$ and $\bar{M}/N$ remain fixed,
$\{K_j\}$ and $\{\gamma_\alpha\}$
are distributed within the intervals [$Q_1$, $Q_2$] and [$B_1$, $B_2$] respectively, with densities $\rho_c(K)$ and $\rho_s(\gamma)$. After some calculations, we obtain the following integral equations for $\rho_c(K)$ and $\rho_s(\gamma)$
\begin{equation}
    \begin{split}
        &2 \pi  \rho _c(K)=-\frac{\sinh (\xi )}{\cosh (\xi )-\cos (K)}-\int_{B_1}^{B_2} \frac{\sinh (\eta ) \rho _s(\gamma )}{\cosh (\eta )-\cos (K-\gamma )} \, d\gamma ,\\
        &2 \pi  \rho _s(\gamma )=-\int_{Q_1}^{Q_2} \frac{\sinh (\eta ) \rho _c(K)}{\cosh (\eta )-\cos (\gamma -K)} \, dK+\int_{B_1}^{B_2} \frac{\sinh (2 \eta ) \rho _s(\lambda )}{\cosh (2 \eta )-\cos (\gamma -\lambda )} \, d\lambda.
    \end{split}
\end{equation}\label{eqrho11}
and the subsidiary conditions
\begin{equation}\label{sbc1}
\int_{Q_1}^{Q_2} \rho _c(K) \, dK=\frac{M}{N},\qquad \int_{B_1}^{B_2} \rho _s(\lambda ) \, d\lambda =\frac{\bar{M}}{N}.
\end{equation}
The energy is 
\begin{equation}
\label{Ex1}
    E=-\int_{Q_1}^{Q_2} \frac{ 2 \sin (K) \sinh (\xi )\rho _c(K)}{\cosh (\xi )-\cos (K)} \, dK.
\end{equation}

Let us consider the half-filling case, i.e., $ M=N$. In this case, $B_2=-B_1=\pi$, the integral equations above can be further simplified via the Fourier transformation
\begin{equation}\label{Half_Filling}
        2 \pi  \rho _c(K)=-\frac{\sinh (\xi )}{\cosh (\xi )-\cos (K)}+\int_{Q_1}^{Q_2} \phi \left(K-K_0\right) \rho _c\left(K_0\right) \, dK_0,
\end{equation}
with
\begin{equation}
            \phi \left(K-K_0\right)=\frac{1}{2}+2\sum _{n=1}^{\infty }\frac{\cos \left(\left(K-K_0\right) n\right)}{1+e^{2 |\eta| n}}.
\end{equation}
\begin{figure}[h]
    \includegraphics[width=0.6\textwidth]{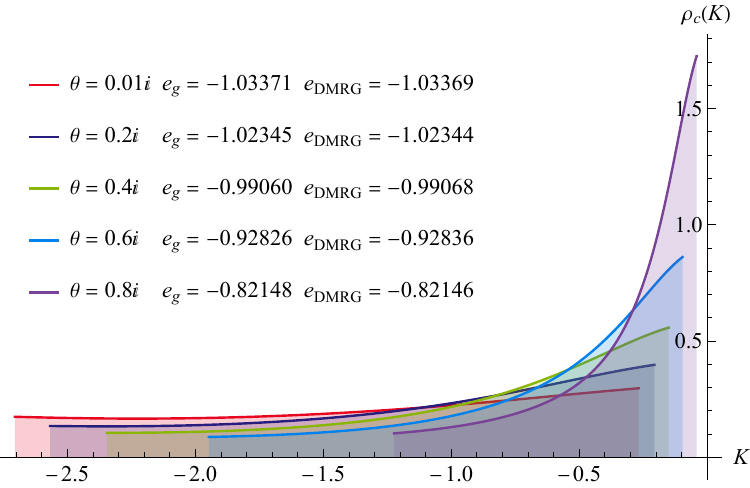} 
    \caption{The numerical result of $\rho(K)$ with $\eta=-0.5$ and  $2\bar{M}/N=1$ for various values of $\theta$.  $e_g=E/N $ and $e_{\text{DMRG}}$ is the result obtained from DMRG.}
    \label{figS1}
\end{figure}

Here, $Q_1$ and $Q_2$ have no explicit expressions and must be determined using equations \eqref{sbc1}, \eqref{Ex1} and \eqref{Half_Filling}.  
For fixed \( \theta \) and \( \eta \), each pair of values \( Q_1, Q_2 \) corresponds to a unique solution $\rho_c(K)$ derived from the integral equation \eqref{Half_Filling}. This solution is associated with a specific filling number (given by \eqref{sbc1}) and an energy value (given by \eqref{Ex1}). The ground state is identified at values of \( Q_1 \) and \( Q_2 \) that corresponds to a filling number of one and minimize energy.  


The numerical results of the ground state $\rho_c(K)$ are shown in Fig.\ref{figS1}.  Due to the odd symmetry of the energy equation \eqref{Ex1}, introducing a non-zero parameter $\theta$ leads to an asymmetric distribution of Bethe roots in the ground state, contrasting with the symmetric distribution observed in the Bariev model \cite{Bariev1}. 
Furthermore, for different parameter values, the integral interval [$Q_1$, $Q_2$] corresponding to the ground state also varies. 
The ground state energy obtained from the above integral equations coincides quite well with the result given by the density matrix renormalization group (DMRG) algorithm calculations. 

\subsection{\(\eta > 0\) and \(0 <\text{Im}[\theta] < 1/h\)}

\begin{figure}[!htbp]
    \centering
    \includegraphics[width=0.7\textwidth]{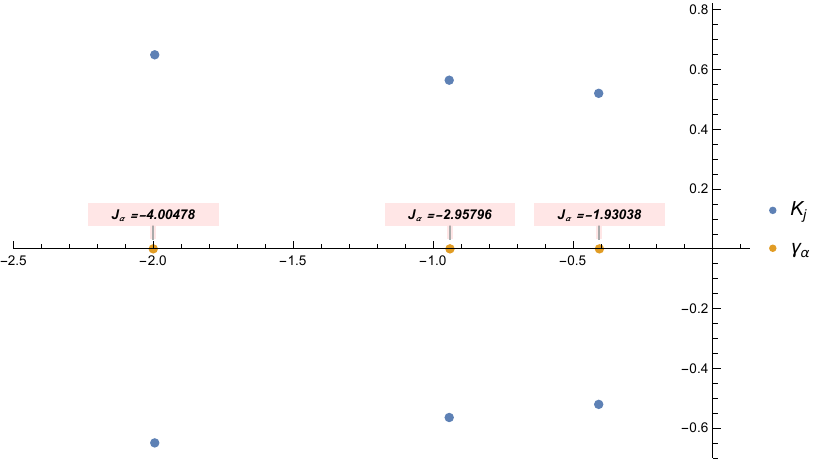}
    \caption{Numerical result of Bethe roots corresponding to the ground state, where \(N=6\), \(M=6\), \(\bar{M}=3\), \(\eta =0.5\) and \(\theta =0.25 i\). Here $K_j$ and $\gamma_\alpha$ are derived from BAEs \eqref{BAEtr1} and $J_\alpha$ is obtained from \eqref{BAEtrr}. Notice that Eq. \eqref{BAEtrr} holds precisely only in the thermodynamic limit. One can observe that the value of $J_\alpha$ exhibits a slight deviation from integers in finite size system. }
    \label{figBie2}
\end{figure}

Here we consider the half-filling case, where the total magnetization of the system's ground state is also zero. For the ground state, $\{K_j\}$ form pairs on the complex plane, while $\{\gamma_\alpha\}$ are all real (see Fig.\ref{figBie2}). In the thermodynamic limit, the Bethe roots satisfy 

\begin{equation}
    K_j=\gamma_\alpha\pm i\eta. 
\end{equation}
Then the BAEs \eqref{BAEtr2} can be simplified to 
\begin{equation}\label{BAEtrr}
    -N \left(\Theta \left(\gamma _{\alpha }+i \eta ,\frac{\xi }{2}\right)+\Theta \left(\gamma _{\alpha }-i \eta ,\frac{\xi }{2}\right)\right)-\sum _{\beta =1}^{\bar{M}} \Theta \left(\gamma _{\alpha }-\gamma _{\beta },\eta \right)=2 \pi  J_{\alpha }, \ \ \alpha=1,...,\bar{M}.
\end{equation}

With $N\to \infty$ and the ratios $M/N$, $\bar{M}/N$ are fixed, the ground state is characterized by the following values of $J_\alpha$ 
\begin{equation}
    J_\alpha=\alpha-\bar{M}-2 \quad \text{for} \quad \alpha=1,...,\bar{M},
\end{equation}
and $\{\gamma_\alpha\}$ fill the interval [$B_1$, $B_2$] uniformly with density $\sigma(\gamma)$. We can obtain the integral equation for $\sigma(\gamma)$ 
\begin{equation}
2 \pi  \sigma (\gamma )=-\frac{\sinh (\xi )}{\cosh (\xi )-\cos (\gamma +i \eta )}-\frac{\sinh (\xi )}{\cosh (\xi )-\cos (\gamma -i \eta )}-\int_{B_1}^{B_2} \frac{\sinh (2 \eta ) \sigma (\lambda )}{\cosh (2 \eta )-\cos (\gamma -\lambda )} \, d\lambda. 
\end{equation}
The values of $B_1$ and $B_2$ are determined by the condition
\begin{equation}
    \int_{B_1}^{B_2} \sigma (\gamma ) \, d\gamma =\frac{\bar{M}}{N}.
\end{equation}
The corresponding energy is 
\begin{equation}
    E=-2 \int_{B_1}^{B_2} \left(\frac{\sinh (\xi ) \sin (\gamma +i \eta )}{\cosh (\xi )-\cos (\gamma +i \eta )}+\frac{\sinh (\xi ) \sin (\gamma -i \eta )}{\cosh (\xi )-\cos (\gamma -i \eta )}\right)\sigma (\gamma )  \, d\gamma.
\end{equation}

\begin{figure}[h]
    \includegraphics[width=0.6\textwidth]{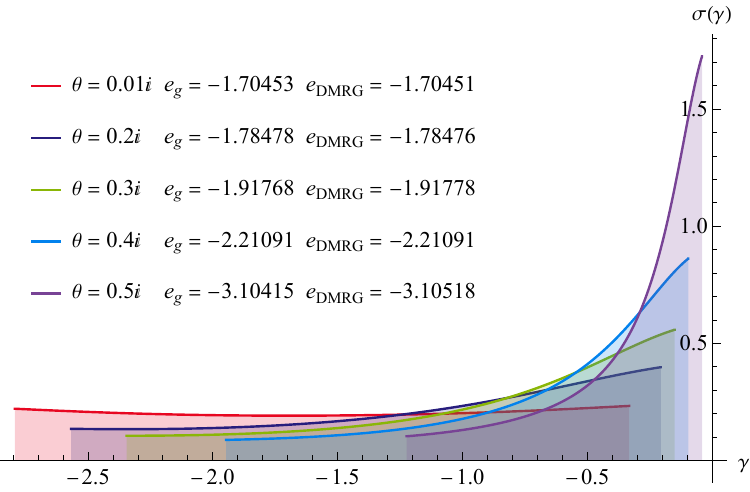} 
    \caption{The numerical result of $\rho(K)$ with $\eta=0.5$ and  $2\bar{M}/N=1$ for various values of $\theta$.  $e_g=E/N $ and $e_{\text{DMRG}}$ is the result obtained from DMRG.}
    \label{figS2}
\end{figure}

The numerical results of the half-filling ground state $\sigma(\gamma)$ are shown in Fig. \ref{figS1}. Similarly, the distribution of Bethe roots of the ground state is asymmetric and the values of $Q_1$ and $Q_2$ change vary with $\theta$. 
We can see that the ground state energy obtained from the above integral equations closely matches the result given by DMRG algorithm calculations.


\section{Single-parameter degenerate model}

As previously mentioned, by setting \(\theta = 0\), the Hamiltonian \eqref{H1} directly reduces to the Bariev model \eqref{HB}.
When $\theta\neq 0$ and $\eta=-\infty$, i.e, $h=0$, we obtain another single-parameter dependent Hamiltonian
\begin{equation}\label{H2}
    \begin{split}
H=&-\sum_{j=1}^N\left\{ \left(c_{j\downarrow}^\dagger c_{j+1\downarrow}+c_{j+1\downarrow}^\dagger c_{j\downarrow}\right)\left( 1+ \left(1-\sqrt{\theta^2+1}\right)n_{j\uparrow}n_{j+1\uparrow}-n_{j\uparrow}\right.\right.
\left.+\left(\sqrt{\theta^2+1}  -1\right) n_{j+1\uparrow}\right)\\
&+\left(c_{j\uparrow}^\dagger c_{j+1\uparrow}+c_{j+1\uparrow}^\dagger c_{j\uparrow}\right)\left( 1+ \left(1-\sqrt{\theta^2+1}\right)n_{j\downarrow}n_{j+1\downarrow}-n_{j+1\downarrow}+\left(\sqrt{\theta^2+1}-1\right) n_{j\downarrow}\right)\\
&\left.-\theta\left( c_{j+1\uparrow}^{\dagger}c_{j\uparrow}c_{j+1\downarrow}^{\dagger}c_{j\downarrow}-c_{j\uparrow}^{\dagger}c_{j+1\uparrow}c_{j\downarrow}^{\dagger}c_{j+1\downarrow}\right)
\right\}.
\end{split}
\end{equation}
Correspondingly, the BAEs and the energy of the system are
\begin{equation}\label{BAEtr3}
    \begin{split}
        \left[\frac{\sin \left(\frac{1}{2} \left(K_j+i \xi \right)\right)}{ \sin \left(\frac{1}{2} \left(K_j-i \xi \right)\right)}\right]^N&=(-1)^{\bar{M}} i^N \prod _{\alpha =1}^{\bar{M}} e^{-i (K_j-\gamma_{\alpha})},\ \ \ j=1,...,M,\\
        \prod _{j=1}^M e^{-i (K_j-\gamma_{\alpha})} &=(-1)^{M-\bar{M}+1}\prod _{\beta =1}^{\bar{M}} e^{-i (\gamma_{\alpha}-\gamma_{\beta})}, \ \ \alpha=1,...,\bar{M},\\   
        E=-\sum_{j=1}^M &\frac{2 \sinh (\xi ) \sin \left(K_j\right)}{\cosh (\xi )-\cos \left(K_j\right)}. 
    \end{split}
\end{equation}
Analogously, to ensure the Hermiticity of the model \eqref{H2}, ${\rm Re}[\theta]$ should be zero and $|\theta|\leq 1$. 

In this case, the Bethe roots $\{K_j\}$ and $\{\gamma_\alpha\}$ corresponding to the ground state are all real. 
In the thermodynamic limit where $N$, $M$, $\bar{M}\to \infty$ and the ratios $M/N$, $\bar{M}/N$ are fixed as finite numbers, the Bethe roots $\{K_j\}$
fill the intervals [$Q_1$, $Q_2$]. The density distribution function $\rho_c(K)$ now satisfies the integral equation
\begin{equation}
    2 \pi  \rho _c(K)=-\frac{\sinh (\xi )}{\cosh (\xi )-\cos (K)}+\frac{1}{2}\int_{Q_1}^{Q_2}  \rho _c\left(K_0\right) \, dK_0, 
\end{equation}
and 
\begin{equation}
    \int_{Q_1}^{Q_2} \rho _c(K) \, dK=\frac{M}{N}.
\end{equation}
The ground state energy is given by 
\begin{equation}
    E=-\int_{Q_1}^{Q_2} \frac{ 2 \sin (K) \sinh (\xi )\rho _c(K)}{\cosh (\xi )-\cos (K)} \, dK.
\end{equation}\label{eqrho3}

\section{Conclusion}

In summary, within the frame of the quantum inverse scattering method, we construct a novel integrable Bariev model that includes more parameters, additional interaction terms, and enriched information of $R$ matrix. Using the Bethe ansatz method, we derive the exact solution of the model. Furthermore, we derive the integral equations for the distribution density of Bethe roots in the ground state and then get the energy. The analysis of the Bethe roots reveals two distinct phase regions for the system's ground state. All our results are confirmed by the numerical calculation. 

Many topics warrant further exploration in the furture. It is straightforward to study the the properties of the system's excited state based on the exact solution. Analogous to the conventional one, the extended Bariev model can also be integrable under open boundary conditions. Constructing and solving the extended Bariev model with open boundaries are other interesting subjects. 

\begin{acknowledgments}
    M. Zheng thanks helpful discussions with A. Kl{\"u}mper and the DMRG calculation support provided by K. Wang. We acknowledge the financial support from National Key R$\&$D Program of China (Grant No.2021YFA1402104),
National Natural Science Foundation of China (Grant Nos. 12074410, 12205235, 12047502, 12247179, 11934015 and 11975183), the Major Basic Research Program of Natural Science of Shaanxi Province
(Grant Nos. 2021JCW-19 and 2017ZDJC-32), and the Strategic Priority Research Program of the Chinese Academy of Sciences (Grant No. XDB33000000).
\end{acknowledgments}

\appendix
\section{Derivation of the Scattering Matrix}
\label{appendix:a}
In this appendix, we present a detailed derivation of the $S$-matrix \eqref{SSSSmatrix}. The eigenstate of the system is given by 
\begin{equation}
        \ket{\Psi} =\sum_{j=1}^M\sum_{\alpha_j=\uparrow,\downarrow}\sum_{x_j=1}^N\psi^{\{\alpha\}} (x_1,\dots,x_M)c_{x_1,\alpha_1}^\dagger \dots c_{x_M,\alpha_M}^\dagger\ket{0},
\end{equation}
where $\psi$ is defined as 
\begin{equation}\label{barievansatzwavefunction}
        \psi^{\{\alpha\}} (x_1,\dots,x_M)=\sum_{p,q}A_p^{\{\alpha\}}(q)\exp\left(i\sum_{j=1}^M k_{p_j}x_{q_j}\right)\theta(x_{q_1}\leq x_{q_2}\leq\dots\leq x_{q_M}).
\end{equation}
Here, $\{\alpha\}=\{\alpha_1,\dots,\alpha_{p_{j}},\alpha_{p_{j+1}},\dots,\alpha_M\}$, $p=\{\dots,p_{j},p_{j+1},\dots \}$, $q=\{\dots,q_{j},q_{j+1},\dots\}$.

Assuming that no two particles occupy the same or adjacent lattice sites, i.e., $x_j\neq x_l$ and $x_{q_j}\neq x_{q_{j+1}}\pm 1$, there is no interactions within the system. Substituting $\ket{\Psi}$ into the eigen-equation $H\ket{\Psi}=E\ket{\Psi}$, we obtain
\begin{equation}
    -2\sum_{j=1}^M\cos k_j \ket{\Psi}=E\ket{\Psi},
\end{equation}
hence $E=-2\sum_{j=1}^M\cos k_j$.

If only two particles are adjacent, $x_{q_j}=x_{q_{j+1}}-1=x$. If the spins of the two particles are aligned ($\uparrow,\uparrow$ or $\downarrow,\downarrow$), there is no interaction between them, and the eigen-equation is automatically satisfied. If the spins are opposite ($\uparrow,\downarrow$ or $\downarrow,\uparrow$), the following neighbor-site equation derived from the eigen-equation equation
\begin{equation}\label{BarievS1}
\begin{split}
       &-h \sqrt{\frac{\theta^2+1}{h^2 \theta^2+1}}\left[\psi^{\{\alpha\}}(...,x_{q_j},x_{q_{j+1}}-1,...)+\psi^{\{\alpha\}}(...,x_{q_j}+1,x_{q_{j+1}},...)\right]\\&-\left[\psi^{\{\alpha\}}(...,x_{q_j},x_{q_{j+1}}+1,...)+\psi^{\{\alpha\}}(...,x_{q_j}-1,x_{q_{j+1}},...)\right]=E\psi^{\{\alpha\}}(...,x_{q_j},x_{q_{j+1}},...).
\end{split}
\end{equation}
Substituting the ansatz (\ref{barievansatzwavefunction}) into (\ref{BarievS1}), we obtain
\begin{equation}\label{BarievSS1}
\begin{split}
       &-h \sqrt{\frac{\theta^2+1}{h^2 \theta^2+1}}\left[\psi^{\{\alpha\}}(...,x,x,...)+\psi^{\{\alpha\}}(...,x+1,x+1,...)\right]\\&-\left[A_p^{\{\alpha\}}(q)e^{2i k_{p_{j+1}}}+A_{p'}^{\{\alpha\}}(q)e^{2i k_{p_j}}+A_p^{\{\alpha\}}(q)e^{i k_{p_{j+1}}-i k_{p_{j}}}+A_{p'}^{\{\alpha\}}(q)e^{2i k_{p_j}}\right]e^{...+ik_{p_j}x+ik_{p_{j+1}}x+...}\\&=E\left(A_p^{\{\alpha\}}(q)e^{i k_{p_{j+1}}}+A_{p'}^{\{\alpha\}}(q)e^{i k_{p_j}}\right)e^{... +ik_{p_j}x+ik_{p_{j+1}}x+ ...}.
\end{split}
\end{equation}
Here, $p'=\{\dots,p_{j+1},p_{j},\dots \}$, $q'=\{\dots,q_{j+1},q_j,\dots\}$.
We get
\begin{equation}\label{ffB1}
    \psi^{\{\alpha\}}(...,x,x,...)=\frac{1}{h}\sqrt{\frac{h^2 \theta^2+1}{\theta^2+1}}\left(A_p^{\{\alpha\}}(q)+A_{p'}^{\{\alpha\}}(q)\right)e^{... +i k_{p_j} x+ i k_{p_{j+1}}x ...},
\end{equation}
indicating that interactions change the form of the wave function when two particles occupy the same site.
Similarly, if $x_{q_j}-1=x_{q_{j+1}}$, substitution into the eigen-equation gives another neighbor-site equation
\begin{equation}\label{BarievS2}
\begin{split}
       &- \sqrt{\frac{\theta^2+1}{h^2 \theta^2+1}}\left[\psi^{\{\alpha\}}(...,x_{q_j}-1,x_{q_{j+1}},...)+\psi^{\{\alpha\}}(...,x_{q_j},x_{q_{j+1}}+1,...)\right]\\&-\left[\psi^{\{\alpha\}}(...,x_{q_j}+1,x_{q_{j+1}},...)+\psi^{\{\alpha\}}(...,x_{q_j},x_{q_{j+1}}-1,...)\right]=E\psi^{\{\alpha\}}(...,x_{q_j},x_{q_{j+1}},...).
\end{split}
\end{equation}
Substituting the ansatz (\ref{barievansatzwavefunction}) and (\ref{ffB1}) into equation (\ref{BarievS2}), simplification yields
\begin{equation}\label{ffB2}
    A_p^{\{\alpha\}} (q)+A_{p'}^{\{\alpha\}} (q)=h\left(A_p^{\{\alpha\}} (q')+A_{p'}^{\{\alpha\}} (q')\right).
\end{equation}

When there are only two particles occupying the same lattice site
$x_{q_j}=x_{q_{j+1}}$, substituting into the eigen-equation, the wave function satisfies the double occupancy equation
\begin{equation}\label{20240628}
\begin{split}
       &- \sqrt{\frac{\theta^2+1}{h^2 \theta^2+1}}\left[\psi(...,x_{q_j}+1,x_{q_{j+1}},...)+\psi(...,x_{q_j},x_{q_{j+1}}-1,...)\right]\\&-h \sqrt{\frac{\theta^2+1}{h^2 \theta^2+1}}\left[\psi(...,x_{q_j},x_{q_{j+1}}+1,...)+\psi(...,x_{q_j}-1,x_{q_{j+1}},...)\right]\\&-\frac{\theta \left(h^2-1\right)}{\theta^2 h^2 +1}\left[\psi(...,x_{q_j}-1,x_{q_{j+1}}-1,...)+\psi(...,x_{q_j}+1,x_{q_{j+1}}+1,...)\right]=E\psi(...,x_{q_j},x_{q_{j+1}},...).
\end{split}
\end{equation}
Substituting (\ref{barievansatzwavefunction}) and \eqref{ffB1} into \eqref{20240628}, we get
\begin{equation}\label{ffB3}
    \begin{split}
        &\left[ \frac{1}{h}\sqrt{\frac{h^2 \theta^2+1}{\theta^2+1}}\left(2\cos{k_{p_j}}+2\cos{k_{p_{j+1}}}+\frac{\theta (h^2-1)}{\theta^2 h^2 +1}\left(e^{i k_{p_j}+i k_{p_{j+1}}}-e^{-i k_{p_j}-i k_{p_{j+1}}}\right)\right) \right.\\
        &\left. -h \sqrt{\frac{\theta^2+1}{h^2 \theta^2+1}} \left(e^{i k_{p_{j+1}}}+e^{-i k_{p_j}}\right) \right] A_p^{\{\alpha\}}(q)\\
        &\left[ \frac{1}{h}\sqrt{\frac{h^2 \theta^2+1}{\theta^2+1}}\left(2\cos{k_{p_j}}+2\cos{k_{p_{j+1}}}+\frac{\theta (h^2-1)}{\theta^2 h^2 +1}\left(e^{i k_{p_j}+i k_{p_{j+1}}}-e^{-i k_{p_j}-i k_{p_{j+1}}}\right)\right) \right.\\
        &\left. -h \sqrt{\frac{\theta^2+1}{h^2 \theta^2+1}} \left(e^{i k_{p_j}}+e^{-i k_{p_{j+1}}}\right) \right] A_{p'}^{\{\alpha\}}(q)\\
        &=\sqrt{\frac{\theta^2+1}{h^2 \theta^2+1}}\left(e^{i k_{p_{j+1}}}+e^{-i k_{p_j}}\right)A_{p}^{\{\alpha\}}(q')+\sqrt{\frac{\theta^2+1}{h^2 \theta^2+1}}\left(e^{i k_{p_j}}+e^{-i k_{p_{j+1}}}\right)A_{p'}^{\{\alpha\}}(q').
    \end{split}
\end{equation}

The fermionic wave function must satisfy exchange antisymmetry, which implies that exchanging the coordinates and spins of two particles will change $\psi$ to $-\psi$, thus
\begin{equation}\label{spinrepresentation}
    A_p^{\{\alpha\}}(q) = -A_p^{\{\alpha'\}}(q').
\end{equation}
here $\{\alpha'\} = \{\dots, \alpha_{p_{j+1}}, \alpha_{p_j}, \dots\}$. 

The scattering matrix $S$ is defined as
\begin{equation}
    A_p^{\{\alpha\}}(q) = S_{p_{j+1}, p_j}(k_{p_{j+1}}, k_{p_j}) A_{p'}^{\{\alpha'\}}(q'),
\end{equation}

Substituting (\ref{spinrepresentation}) into (\ref{ffB2}) and (\ref{ffB3}), we obtain
\begin{equation}\label{ffB4}
    \begin{split}
         &A_p^{\uparrow,\downarrow} (q) + h A_p^{\downarrow,\uparrow} (q) = A_{p'}^{\downarrow,\uparrow} (q') + h A_{p'}^{\uparrow,\downarrow} (q'),\\
         &\left[ \frac{1}{h}\sqrt{\frac{h^2 \theta^2+1}{\theta^2+1}}\left(2\cos{k_{p_j}}+2\cos{k_{p_{j+1}}}+\frac{\theta (h^2-1)}{\theta^2 h^2 +1}\left(e^{i k_{p_j}+i k_{p_{j+1}}}-e^{-i k_{p_j}-i k_{p_{j+1}}}\right)\right) \right.\\
         &\left. -h \sqrt{\frac{\theta^2+1}{h^2 \theta^2+1}} \left(e^{i k_{p_{j+1}}}+e^{-i k_{p_j}}\right) \right] A_p^{\uparrow,\downarrow}(q) + \sqrt{\frac{\theta^2+1}{h^2 \theta^2+1}}\left(e^{i k_{p_{j+1}}}+e^{-i k_{p_j}}\right)A_p^{\downarrow,\uparrow}(q)\\
         =&\left[ \frac{1}{h}\sqrt{\frac{h^2 \theta^2+1}{\theta^2+1}}\left(2\cos{k_{p_j}}+2\cos{k_{p_{j+1}}}+\frac{\theta (h^2-1)}{\theta^2 h^2 +1}\left(e^{i k_{p_j}+i k_{p_{j+1}}}-e^{-i k_{p_j}-i k_{p_{j+1}}}\right)\right) \right.\\
         &\left. -h \sqrt{\frac{\theta^2+1}{h^2 \theta^2+1}} \left(e^{i k_{p_j}}+e^{-i k_{p_{j+1}}}\right) \right] A_{p'}^{\downarrow,\uparrow}(q') + \sqrt{\frac{\theta^2+1}{h^2 \theta^2+1}}\left(e^{i k_{p_j}}+e^{-i k_{p_{j+1}}}\right)A_{p'}^{\uparrow,\downarrow}(q').
    \end{split}
\end{equation}

Calculations show
\begin{equation}
    \begin{split}
        &A_p^{\uparrow,\downarrow}(q)=\frac{e^{\eta } \left(\theta ^2+1\right) \left(e^{i k_{p_{j}}}-e^{i k_{p_{j+1}}}\right)A_{p'}^{\uparrow,\downarrow} (q')+\left(e^{2 \eta }-1\right) \left(e^{i k_{p_{j}}}+\theta \right) \left(\theta  e^{i k_{p_{j+1}}}-1\right)A_{p'}^{\downarrow,\uparrow} (q')}{\theta ^2 \left(e^{2 \eta +i k_{p_{j}}}-e^{i k_{p_{j+1}}}\right)+\left(e^{2 \eta }-1\right) \theta  \left(-1+e^{i \left(k_{p_{j}}+k_{p_{j+1}}\right)}\right)+\left(e^{i k_{p_{j}}}-e^{2 \eta +i k_{p_{j+1}}}\right)}\\
        &A_p^{\downarrow,\uparrow}(q)=\frac{\left(e^{2 \eta }-1\right) \left(e^{i k_{p_{j+1}}}+\theta \right) \left(\theta  e^{i k_{p_{j}}}-1\right)A_{p'}^{\uparrow,\downarrow} (q')+e^{\eta } \left(\theta ^2+1\right) \left(e^{i k_{p_{j}}}-e^{i k_{p_{j+1}}}\right)A_{p'}^{\downarrow,\uparrow} (q')}{\theta ^2 \left(e^{2 \eta +i k_{p_{j}}}-e^{i k_{p_{j+1}}}\right)+\left(e^{2 \eta }-1\right) \theta  \left(-1+e^{i \left(k_{p_{j}}+k_{p_{j+1}}\right)}\right)+\left(e^{i k_{p_{j}}}-e^{2 \eta +i k_{p_{j+1}}}\right)}.
    \end{split}
\end{equation}
In the case of $\alpha_{p_j}=\alpha_{p_{j+1}}$, there is no interaction and one can get $S_{11}$ and $S_{44} = 1$.
Finally, we obtain the scattering matrix \eqref{SSSSmatrix}.

\section{Solution of the second eigenvalue problem}
\label{appendix:b}
The $S$ matrix satisfies the Yang-Baxter equation 
\begin{equation}
    \mathcal{S}_{1,2}(\lambda_1,\lambda_2)\mathcal{S}_{1,3}(\lambda_1,\lambda_3)\mathcal{S}_{2,3}(\lambda_2,\lambda_3)=\mathcal{S}_{2,3}(\lambda_2,\lambda_3)\mathcal{S}_{1,3}(\lambda_1,\lambda_3)\mathcal{S}_{1,2}(\lambda_1,\lambda_2). 
\end{equation}
Then we can define a new inhomogeneous monodromy matrix and transfer matrix generated from the $S$ matrix
\begin{equation}
    \begin{split}
        \widetilde{T}_0 (u)&=\mathcal{S}_{0,M}(u,\lambda_M) \mathcal{S}_{0,M-1} (u,\lambda_{M-1})\dots \mathcal{S}_{0,2} (u,\lambda_2) \mathcal{S}_{0,1} (u,\lambda_1)\\
        &=\begin{pmatrix}
            A(u)&B(u)\\
            C(u)&D(u)
            \end{pmatrix},\\
        \widetilde{t}(u)&=tr_0 \{\widetilde{T}_0 (u)\}, 
     \end{split}
\end{equation}
where $A (u)$, $B (u)$, $C (u)$, $D (u)$ are operators depending on the spectral parameter $u$.
We can see that $\widetilde{t}(\lambda_j)$ can be expressed as
\begin{equation}
    \widetilde{t}(\lambda_j)=\mathcal{S}_{j,j+1}(\lambda_j,\lambda_{j+1})\mathcal{S}_{j,j+2}(\lambda_j,\lambda_{j+2})\dots \mathcal{S}_{j,M}(\lambda_j,\lambda_M)\mathcal{S}_{j,1}(\lambda_j,\lambda_1)\dots \mathcal{S}_{j,j-1}(\lambda_j,\lambda_{j-1}), 
\end{equation}
which is exactly the left-hand side of (\ref{SSe}). 

From the algebraic Bethe ansatz method, 
the commutation relations among the elements of the monodromy matrix can be obtained 
\begin{equation}
    \begin{split}\label{commutationrelation}
  [A(u),A(v)]&=[B(u),B(v)]=[C(u),C(v)]=[D(u),D(v)]=0,\\
 A(u) B(v) = &  \frac{\sin \left(\frac{u-v}{2}-i \eta \right)}{\sin \left(\frac{u-v}{2}\right)}   B(v)A(u)+ \frac{i \sinh (\eta ) e^{-\frac{i}{2} ( u- v)}}{\sin \left(\frac{u-v}{2}\right)} B(u)A(v),\\
D(u) B(v)=& \frac{\sin \left(\frac{u-v}{2}+i \eta \right)}{\sin \left(\frac{u-v}{2}\right)} B(v)D(u)-\frac{i \sinh (\eta ) e^{-\frac{i}{2} ( u- v)}}{\sin \left(\frac{u-v}{2}\right)} B(u)D(v).
\end{split}
\end{equation}
The vacuum state of the system can be defined as 
\begin{equation}
\ket{\varphi}=\bigotimes_{n=1}^N\binom{1}{0}.
\end{equation} 
 We have 
\begin{equation}
\begin{split}
    A(u)\ket{\varphi}&=a(u)\ket{\varphi}=\ket{\varphi},\\
    D(u)\ket{\varphi}&=d(u)\ket{\varphi}=\prod_{l=1}^M \frac{\sin \left(\frac{1}{2} \left(u-\lambda _l\right)\right)}{\sin \left(\frac{1}{2} \left(u-\lambda _l\right)+i \eta \right)}\ket{\varphi},\\
    C(u)\ket{\varphi}&=0.    
\end{split}
\end{equation}
Then the operator $B(u)$ can be treated as the spin flipping operator and used to construct the Bethe state
\begin{equation}
    \ket{\mu_{1},\mu_2,\dots,\mu_{\bar{M}}}=\prod_{\alpha=1}^{\bar{M}}B(\mu_\alpha)\ket{\varphi},
\end{equation}
where $\bar{M}\leq M$ is a new quantum number  and $\mu_\alpha$  can be considered as  spin rapidities. 
With the help of commutation relations~\eqref{commutationrelation}, we can get
\begin{equation}
    \begin{split}
        \widetilde{t}(u)\ket{\varphi}=&\left( A(u)+D(u)\right)\ket{\varphi}\\
    =&\Lambda(u)\ket{\mu_{1},\mu_2,\dots,\mu_{\bar{M}}}\\
    &-\sum_{\alpha=1}^{\bar{M}}\Lambda_j (u)B(\mu_1)B(\mu_2)\dots B(\mu_{\alpha-1})B(u)B(\mu_{\alpha+1})\dots B(\mu_{\bar{M}})\ket{\varphi}, 
    \end{split}
\end{equation}
where
\begin{equation}
    \begin{split}       
        \Lambda(u)=&a(u)\prod_{\alpha=1}^{\bar{M}}\frac{\sin \left(\frac{1}{2} \left(u-\mu _{\alpha }\right)-i \eta \right)}{\sin \left(\frac{1}{2} \left(u-\mu _{\alpha }\right)\right)}+d(u)\prod_{\alpha=1}^{\bar{M}} \frac{\sin \left(\frac{1}{2} \left(u-\mu _{\alpha }\right)+i \eta \right)}{\sin \left(\frac{1}{2} \left(u-\mu _{\alpha }\right)\right)},\\
        \Lambda_j (u)=&\frac{i \sinh (\eta ) e^{-\frac{i}{2} ( u- \mu_\alpha)}}{\sin \left(\frac{u-\mu_\alpha}{2}\right)}\left(a(\mu_{\alpha})\prod_{\beta \neq \alpha}^{\bar{M}}\frac{\sin \left(\frac{1}{2} \left(\mu _{\alpha }-\mu _{\beta }\right)-i \eta \right)}{\sin \left(\frac{1}{2} \left(\mu _{\alpha }-\mu _{\beta }\right)\right)} -d(\mu_{\alpha}) \prod_{\beta \neq \alpha}^{\bar{M}} \frac{\sin \left(\frac{1}{2} \left(\mu _{\alpha }-\mu _{\beta }\right)+i \eta \right)}{\sin \left(\frac{1}{2} \left(\mu _{\alpha }-\mu _{\beta }\right)\right)}\right) . 
    \end{split}
\end{equation}
To ensure the Bethe state is an eigenstate of the transfer matrix, $\Lambda_j$ should be zero. Combined with equation~\eqref{SSe} and perform parameter transformation $\theta=i e^{\xi}$, $K_j=\lambda_j+\frac{\pi}{2}$ and $\gamma _{\alpha }= \mu_{\alpha }-i \eta -\frac{\pi }{2}$, 
we can get the BAEs 
\begin{equation}
    \begin{split}
        \left[\frac{\sin \left(\frac{1}{2} \left(K_j+i \xi \right)\right)}{ \sin \left(\frac{1}{2} \left(K_j-i \xi \right)\right)}\right]^N=&i^N\prod _{\alpha =1}^{\bar{M}} \frac{\sin \left(\frac{1}{2} \left(K_j-\gamma _{\alpha }-i \eta\right)\right)}{\sin \left(\frac{1}{2} \left(K_j-\gamma _{\alpha }+i \eta \right)\right)},\\
        \prod _{j=1}^M \frac{\sin \left(\frac{1}{2} \left(K_j-\gamma_{\alpha }-i \eta \right)\right)}{\sin \left(\frac{1}{2} \left(K_j-\gamma_{\alpha }+i \eta \right)\right)}&=-\prod _{\beta =1}^{\bar{M}} \frac{\sin \left(\frac{1}{2} \left(\gamma _{\alpha }-\gamma _{\beta }\right)+i \eta \right)}{\sin \left(\frac{1}{2} \left(\gamma _{\alpha }-\gamma _{\beta }\right)-i \eta \right)}.   
    \end{split}
    \end{equation}
The corresponding eigenvalue is 
\begin{equation}
    E=-2\sum_{j=1}^M \cos{k_j}=-\sum_{j=1}^M \frac{2 \sinh (\xi ) \sin \left(K_j\right)}{\cosh (\xi )-\cos \left(K_j\right)}. 
\end{equation}

\end{document}